%% file: entanglv2.tex
\documentclass[12pt]{article}
\usepackage{graphicx}
\usepackage{psfrag,amssymb}
\usepackage{epic,epsfig}
\usepackage[dvips]{color}

\newcommand{\eq}{\begin{equation}}
\newcommand{\en}{\end{equation}}
\newcommand{\ear}{\begin{eqnarray}}
\newcommand{\rae}{\end{eqnarray}}
\newcommand{\Z}{\mathbb{Z}}
\newcommand{\C}{\mathbb{C}}

\newcommand{\os}{\mathbb{O}}
\newcommand{\R}{{\cal R}}

\newcommand{\D}{{\cal D}}

\newcommand{\bra}{\langle}
\newcommand{\ket}{\rangle}

\newcommand{\tr}{{\rm tr}\,}

\begin{document}
\renewcommand{\theequation}{\thesection.\arabic{equation}}

\begin{titlepage}

\centerline{\Large{Entanglement Entropy and Twist Fields}}
\bigskip\bigskip
\bigskip\bigskip
\centerline{\it Michele Caraglio, Ferdinando Gliozzi }
\medskip
\centerline{
Dipartimento di Fisica Teorica, Universit\`a di Torino  and}
\centerline{ INFN,
Sezione di Torino, via P. Giuria, 1, I-10125 Torino, Italy}
\medskip
\medskip
\medskip
\bigskip\bigskip

\begin{abstract}
 The entanglement entropy of a subsystem $A$ of 
a quantum system is expressed, in the replica approach, through analytic 
continuation with 
respect to $n$ of the trace of the $n-$th power of the reduced density 
matrix. This trace can be thought of as the vacuum 
expectation value of a suitable 
observable in a system made with $n$ independent copies of the original 
system. We use this property to numerically evaluate it in 
some two-dimensional critical systems, where it can be compared with the 
results of Calabrese and Cardy, who wrote the same quantity in terms of 
correlation functions of twist fields of a conformal field theory. 
Although the two calculations
match perfectly even in finite systems when the system $A$ consists  
of a single interval, they  
disagree whenever the subsystem $A$ is composed of more than one connected 
part. The reasons of this disagreement are explained.
\end{abstract}

\end{titlepage}
 \baselineskip=18pt \setcounter{footnote}{0}

\section{Introduction}

In a quantum system, performing a local measurement may instantaneously affect 
local measurements far away. This is a manifestation of the quantum 
entanglement. A useful measure of this property  
in extended quantum systems with many degrees of freedom is the von Neumann 
or entanglement entropy: one considers a pure quantum state
$\vert\Psi\ket$ (typically the ground state) that can be subdivided into two subsystems $A$ and $B$ and constructs the reduced density matrix
\eq 
\rho_A=\tr_B\,\vert\Psi\ket\bra\Psi\vert
\en 
by tracing over the degrees of freedom of $B$. The von Neumann or entanglement
entropy is defined to be
\eq
S_A\equiv-\tr\rho_A\ln \rho_A=-\tr\rho_B\ln \rho_B\equiv S_B~.
\label{eentropy}
\en

Of particular interest is the case in which the two subsystems $A$ and $B$ correspond to two connected regions (the ``outside'' and the ``inside'') in the space-time, where it was argued that the entanglement (or geometric) entropy  is deeply related to the physics of the black holes\cite{'t Hooft:1984re,sr,ge}.
An important element of the present understanding is its holographic 
interpretation \cite{Ryu:2006bv}. This point of view seems to indicate a purely
geometric way of computing  the entanglement entropy in strongly coupled conformal field theories \cite{Emparan:2006ni}.

The entanglement entropy has been also extensively studied in low dimensional quantum systems as a new tool to investigate the nature of quantum criticality
\cite{hlw,vid,ch,cc,cp}.
Several different calculations based on the conformal field theory (CFT) 
describing the universal properties of the quantum phase transition describing 1+1 dimensional systems, like quantum spin chains, have shown that the entropy grows 
logarithmically with the size $\ell$ of the subsystem $A$ as 
\cite{hlw,vid,ch,cc}
\eq
S_A=\frac c3 \ln(\ell/a)~,
\label{esc}
\en
where $c$ is the conformal anomaly and $a$ an ultraviolet cutoff.

The trace of the $n$-th power of the reduced density matrix, 
which could be identified with the Tsallis entropy 
  \cite{tsallis}, plays a major role in the replica approach to entanglement 
entropy \cite{cc}, yielding
\eq
S_A=-\lim_{n\to1}\frac\partial{\partial n}\tr\rho_A^n~.
\en  

  Our goal in  this paper is to describe a simple method to directly measure 
$\tr\rho_A^n$  in whatever local lattice field theory in any space-time 
dimension, based on the 
observation that this quantity can be expressed, as we shall explain below, 
as the vacuum expectation value of a suitable observable  defined in a 
system made with $n$ independent copies of the original quantum system.

We apply our method to check various consequences of the replica approach 
in some $(1+1)$-dimensional 
critical quantum system described by a relativistic conformal field  theory 
(CFT), including a system with $c<0$, where the 
entanglement entropy appears to be negative, finding also in that case complete agreement: evidently in those 
non-unitary systems the mixed state obtained by tracing over the degrees  
of freedom of $B$ is more ordered than the pure state $\vert\Psi\ket$.

When the  subsystem $A$ consists  of $N$ disjoint intervals
$A\equiv$ $(u_1,v_1)\dots (u_N,v_N)$ the quantity $\tr\rho_A^n$ is 
proportional,
at criticality, to the $n-$th power of the correlation function of $2N$ 
local primary operators of complex scaling dimensions
\eq
\Delta_n=\bar\Delta_n=\frac c{24}(1-\frac1{n^2})
\label{delta}
\en 
sitting on the end points of the $N$ intervals:
\eq
\tr \rho_A^n\propto \bra\prod_{j=1}^{N}\Phi_n^-(u_j,\bar{u}_j)\Phi_n^+(v_j,\bar{v}_j)\ket^n~.
\label{rhobp}
\en 
Notice that these primary operators do not belong to the Kac table. 
They can be considered as a special kind of twist fields, called 
branch-point twist fields \cite{ccd}, because they are naturally related 
to the branch points in the $n$-sheeted Riemann surface where  
the system is defined in its Euclidean functional integral formulation.
These primary operators also carry  a conserved charge 
related to the orientation of the intervals.
 
The proof of the above statements relies (among other things) 
on the clever observation \cite{cc} that the vacuum expectation value of 
the stress tensor $T(z)$ near a branch point has the same functional
dependence as the one generated by a primary operator with scaling dimensions
$\Delta_n$. Since the same branch point is present in each sheet of the 
Riemann surface, it is easy to conclude that $\tr \rho_A^n$ has the form
(\ref{rhobp}).

By making a further assumption  we shall discuss in the next Section, 
the authors of \cite{cc} proposed an explicit functional form
\footnote{There is a typo in this formula in \cite{cc}. We thank P.Calabrese 
for pointing this out to us. }
\eq
\tr \rho_A^n\propto \left\vert\frac{\prod_{j<k\le N}(u_k-u_j)(v_k-v_j)}
{\prod_{j, k\le N} (v_k-u_j)}\right\vert^{4n\Delta_n}~,
\label{wrong}
\en
Note that the points of type $u$ and of type $v$, associated respectively  
with $\Phi^-_n$ and $\Phi^+_n$, are treated in a different way. This is 
justified only when the intervals are intrinsically oriented, which is not the 
case  when $n=2$, where all the branch points are on the  same footing 
and $\Phi^-_2(w)\equiv \Phi^+_2(w)$. 
We shall discuss below the $N=n=2$ case in detail. 
When $n$ is an odd number 
and $c>0$ we shall  argue, and corroborate with numerical experiments, 
that whenever 
two twist fields of the same charge approach each other the above quantity 
develops an ultraviolet divergence instead of vanishing as 
 (\ref{wrong}) would require. 

The origin of such a 
discrepancy can be traced  to the  assumption that the only singularities 
of $T(z)$ are the double poles at the branch points of the 
multi-sheeted Riemann 
surface, which is not the case of the conformal mappings $w=f(z)$
considered in \cite{cc}. This point will be further discussed  in the next
Section.

The contents of this paper are as follows. In the next Section we summarise the main results of the replica approach to entanglement entropy in CFT and discuss
some difficulties in the use of the conformal mappings when the number of branch points is larger than two. In the following Section we describe our 
method of evaluating $\tr \rho^n_A$ as the vacuum expectation value of a 
suitable observable; as a byproduct  we uncover a special quantum 
symmetry, which is exact only at zero temperature,  yielding at once the 
important identity $\tr \rho_A^n=\tr\rho_B^n$ and other useful relationships. 
In Section \ref{Potts} we implement the method in some specific 
two-dimensional systems, namely the Potts models: they can be easily 
simulated at their critical point and 
the corresponding  CFT is known. We finish with some conclusions.           
\setcounter{equation}{0}
\section{Conformal maps and twist fields}
\label{cft}
Translational and rotational invariance of a CFT on the complex $z-$plane $\C$
imply $\bra T(z)\ket_\C=0$. Under a conformal mapping $w\to z=f(w)$ the 
stress tensor $T$ transforms as
\eq
T(w)=\left(\frac{df}{dw}\right)^2\,T(z)+\frac c{12}\{f,w\}
\label{transf}
\en
where
\eq
\{f,w\}=\frac{f'''}{f'}-\frac32 \left(\frac{f''}{f'}\right)^2
\label{swz}
\en
is called Schwarzian derivative. Comparing the vacuum expectation value of 
both sides of (\ref{transf}) yields
\eq
\bra T(w)\ket_\R=\frac c{12}\{f,w\}
\label{twz}
\en
where $\R$ is the Riemann surface associated with $f(w)$. 

The  identity
$  \{f,w\}=-2\sqrt{f'}\frac{d^2}{dw^2}\frac1{\sqrt{f'}}$ implies at once that
$\{f,w\}\equiv0$ if and only if $f(w)$ is a linear fractional transformation,
which is the only conformal mapping of the whole $\C$ onto itself. In any other
case $\bra T(w)\ket\not\equiv0$ and some symmetry of the original system is 
lost. In particular, using again the above-mentioned identity, it is easy to 
see that the only conformal mappings conserving translational invariance
( hence $\bra T(w)\ket_\R=const.$) are the exponential mappings, 
which can be written in a suitable basis in the form
\eq
z=\exp( {2\pi}\frac wL)~,
\label{exp}
\en
which maps the cylinder composed of the infinite strip $0\le \Im m\,w\equiv 
y<L$ with periodic boundary conditions into the whole $z$ plane 
excepted the origin.
As a parenthetic remark, we note that this exponential mapping is the key 
ingredient to find  universal finite size effects of two-dimensional 
field theories at criticality \cite{car}. In any other conformal mapping,
being $\bra T(w)\ket$ a non-constant analytic function, it has to be 
singular somewhere. By dimensional reasons, the isolated singularities 
of $\bra T(w)\ket$ are double poles.  

As a simple illustrative example, apply  (\ref{twz})  to   
the mapping $z=f(w)=w^{\frac1n}$. It yields at once
\eq
\bra T(w)\ket_{\R_n}=\frac c{24}\frac{(1-1/n^2)}{w^2}~,
\en 
where $\R_n$ is now a $n-$sheeted Riemann surface  branched over 
$0$ and   $\infty$. We can now first displace these points 
with a linear fractional transformation
$g(w)=\frac{w-u}{w-v}$ and then consider the composed transformation 
\eq
z(w)=f(g(w))=\left(\frac{w-u}{w-v}\right)^{\frac1n}~.
\en
 Being $g$  a linear fractional transformation, 
we may take advantage of the remarkable identity  
$\{f(g),w\}={g'}^{\,2}\,\{f,g\} $ to get
 \eq
\bra T(w)\ket_{\R_n}=c\frac{(1-1/n^2)}{24}\frac{(u-v)^2}{(w-u)^2(w-v)^2}~.
\en 
 
As first pointed out in \cite{cc}, this expression coincides, up 
to a normalising constant,  with the three-point function 
$\bra T(w)\Phi_n(u,\bar{u})\Phi_n(v,\bar{v})\ket_\C$ where 
$\Phi_n(w,\bar{w})$ is a primary operator, called branch-point twist field
\cite{ccd}, with complex scaling dimensions 
$\Delta_n=\bar\Delta_n=c\frac{1-1/n^2}{24}$ associated with the branch points
$w=u$ and $w=v$ of $\R_n$, as already anticipated in the Introduction. 
Combining such an observation with the fact that  the surface 
constructed with the replica method has the same intrinsic geometric 
properties as $\R_n$ Calabrese and Cardy \cite{cc} were able to argue that 
\eq
\tr \rho_A^n\propto \bra \Phi_n^-(u,\bar{u})\Phi_n^+(v,\bar{v})\ket_\C^n=c_n
\left\vert\frac a{ u-v}\right\vert^{\frac c6(n-1/n)}~,
\label{uv}
\en 
where $a$ is an UV cutoff and $A$ is the interval joining the branch points
$u$ and $v$.  

To make contact with numerical simulations it is useful to rewrite this 
correlator in a finite geometry using the transformation law of the primary 
operators under a suitable conformal mapping $ w=h(\zeta)$ :
\eq
\bra\Phi_n^-(\zeta_1,\bar{\zeta}_1)\Phi_n^+(\zeta_2,\bar{\zeta}_2)\ket_\R=
\vert w'(\zeta_1)w'(\zeta_2)\vert^{2\Delta_n}
  \bra\Phi_n^-(u,\bar{u})\Phi_n^+(v,\bar{v})\ket_\C~,
\en   
where $\zeta_1$ and $\zeta_2$ are the inverse images of $u$ and $v$ 
and $\R$ is the Riemann surface associated with $h$.
Using the exponential mapping (\ref{exp}) and arranging for 
instance the branch points to lie on the imaginary axis  of the cylinder $\zeta=x+i\,y$, one 
finally finds \cite{cc}
\eq
\tr\rho_A^n=c_n\bra\Phi_n^-(y_1)\Phi_n^+(y_2)\ket_{\rm cyl.}^n=
c_n\left(\frac{\pi/L}{\sin{(\pi\ell/L)}}\right)^{\frac c6(n-1/n)}~,
\label{sine}
\en
with $\ell=\vert y_2-y_1\vert$.

We checked this formula with numerical experiments in two different
 critical systems, one, the Ising model, 
with positive conformal anomaly $(c=\frac12)$ and the other, a $Q$-state 
Potts model with $Q<1$, corresponding to a conformal theory with a negative 
conformal anomaly $(c=-\frac{11}{14})$. In either case we studied the 
dependence on $n$, $L$ and $\ell$, finding perfect agreement 
with (\ref{sine}) provided the distances
of the branch points and the sizes of the lattice are large enough with 
respect to the lattice spacing (see Section \ref{Potts} for more details).

\subsection{General case}
When  A consists of more than one disjoint interval,
i.e. $A\equiv (u_1,v_1)\dots (u_N,v_N)$, it is natural to assume that 
the Tsallis entropy $\tr \rho_A^n$ be proportional to the
$n-$th power of the  correlation function
of the twist fields associated with the $2N$ branch 
points (see Eq.(\ref{rhobp})).

The  Ward identities of the conformal invariance  tell us that 
the only singularities of $\bra T(w)\Phi_n^-(u_1,\bar{u}_1)\dots 
\Phi_n^+(v_N,\bar{v}_N)\ket $ 
are the double poles located at $u_j,v_j$ $(j=1\dots N)$. Ca\-la\-bre\-se and 
Cardy proposed the Ansatz
\eq  
\frac{\bra T(w)\Phi_n^-(u_1,\bar{u}_1)\dots \Phi_n^+(v_N,\bar{v}_N)\ket_\C}
{\bra \Phi_n^-(u_1,\bar{u}_1)\dots \Phi_n^+(v_N,\bar{v}_N)\ket_C}\equiv
\bra T(w)\ket_{\R_{n\,N}}=\frac c{24} \{f_{n\,N},w\}~,
\label{twrong}
\en
with
\eq
z=f_{n\,N}(w)=\prod_{j=1}^N\left(\frac{w-u_j}{w-v_j}\right)^{\frac1n}~;
~~u_j\not=u_k~,~v_j\not=v_k~\forall ~j\not=k~.
\label{fwrong}
\en

\begin{figure}[tb]
\begin{center}
\mbox{~\epsfig{file=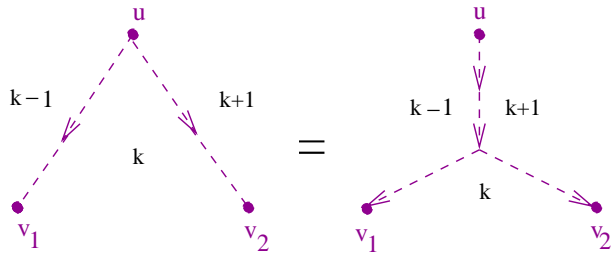,width=8.cm}}
\vskip .2 cm
\caption{Three branch points generated by the fusion of two oriented intervals
$(u_1,v_1)\,(u_2,v_2)$ in the limit $\vert u_1-u_2\vert\to 0$. A general symmetry of the replica approach described in Section \ref{usy} allows to deform 
the intervals as indicated on the right.}
\label{Figure:0}
\end{center}
\end{figure}

The condition $u_j\not=u_k\,,~v_j\not=v_k~\forall ~j\not=k$, which is an important ingredient to derive (\ref{wrong}), has no justification in the
 process of sewing  the $n$ replicas, as there is no geometrical obstruction in letting two branch points of the same type, say $u$, coincide, like 
in Figure  \ref{Figure:0}. 

The equality illustrated in Figure \ref{Figure:0} comes from the fact that, 
in the physical systems we are studying, a cut joining two branch 
points may be continuously deformed, as we shall discuss in detail in the next
Section. Note that the cut associated with the fused branch point sews together
the sheets in the order $k\to k+2$. Clearly the order in which the replicas
are sewn together is immaterial, provided that all the $n$ replicas are 
cyclically connected. On the other hand, if $n$ is an odd number, the 
permutation $k\to k+2$ covers once all the $n$ replicas. Therefore, 
according to the general argument of \cite{cc}, the complex scaling 
dimensions of the 
twist field $\Phi^{+2}_n$ arising from the fusion of  two twist fields with 
equal charge are expected to be exactly the same, i.e. 
$\Delta_n=\bar{\Delta}_n=\frac c{24}(1-\frac1{n^2})$.
We can take advantage of this remarkable property to write down the 
operator product expansion (OPE)
\eq
\Phi_n^+(0,0)\,\Phi_n^+(w,\bar{w})=
\frac1{\vert w\vert^{2\Delta_n}}\Phi^{+2}_n(0,0)+\dots 
\label{ope}
\en
where $n$ is an odd number. The dots indicate less singular terms. 

Similarly  (\ref{uv}) yields
\eq
\Phi_n^+(0,0)\,\Phi_n^-(w,\bar{w})=
\frac{c_n}{\vert w\vert^{4\Delta_n}}+\dots 
\label{opm}
\en
Comparing  (\ref{ope}) and (\ref{opm}) shows
that two branch points of the same type develop, as the interdistance decreases, an UV singularity with an exponent which is exactly half the exponent 
associated with a pair of branch points of opposite type, while
(\ref{wrong}) would predict a zero. A numerical check of this behaviour is 
illustrated in Figure \ref{Figure:6}.  

In the limit configuration depicted in 
Figure \ref{Figure:0}, $\tr\rho_A^n$  becomes of course proportional to 
the $n-$th power of the three-point function
\eq
\tr\rho^n(u,v_1,v_2)\propto\bra\Phi^{-2}_n(u,\bar{u}),\Phi^+_n(v_1,\bar{v}_1),
\Phi^+_n(v_2,\bar{v}_2)\ket^n_\C~,
\en
therefore conformal invariance completely fixes its functional dependence:
\eq
\tr\rho^n(u,v_1,v_2)\propto
\frac{1}{\vert(v_1-v_2)( v_2-u)(u-v_1)\vert^{2n\Delta_n} }~,
\label{three}
\en    
which is valid for any odd $n$. A numerical check of the scaling behaviour 
of this quantity for $n=3$  can be found in Figure \ref{Figure:5}. 

Going back to (\ref{twrong}) and (\ref{fwrong}), note that they are 
based on two distinct assumptions: 
\begin{itemize}
\item[\emph{i)}] $\bra T(w)\ket_{R_{n\,N}}$  can be written 
as the Schwarzian derivative of a suitable function $f_{n\,N}$;  
\item[\emph{ii)}] this function is given by (\ref{fwrong}). 
\end{itemize}
It is easy to prove that the latter assumption is not correct in the case 
of multi-intervals  
because the function $f_{n,N>1}(w)$  does not fulfil the condition that the 
only singularities of its Schwarzian derivative $\{f_{n,N},w\}$ are the 
double poles at $u_j$ and $v_j$, 
as  conformal Ward identities require. 
A simple way to see it, 
is to put $n=1$ in (\ref{fwrong}), thereby eliminating  all the branch 
points and the associated twist fields. If the only singularities of
(\ref{twrong}) were the double poles at $u_j,v_j$ ($j=1\dots N$) one should
have $\{f_{1\,N},w\}=0$. This would imply, as emphasised at the beginning 
of this Section,  $f_{1\,N}(w)$  to be a linear fractional 
transformation, which evidently is not the case when $N>1$.
Actually we find, in the $N=2$ case,
 \eq
 \{f_{1\,2},w\}=-\frac32\frac1{(w-r_1)^2}-\frac32\frac1{(w-r_2)^2}
+\frac3{r_1-r_2}\left(\frac1{w-r_1}-\frac1{w-r_2}\right)~, 
\en
where $r_1$ and $r_2$ are the two roots of the equation
$\frac{d\,f_{1\,2}}{d\,w}=0$, namely
\eq
r_{1,2}=\frac{v_1v_2-u_1u_2\pm\sqrt{(u_1-v_1)(v_1-u_2)(u_2-v_2)(v_2-u_1)}}
{u_1-v_1+u_2-v_2}~.
\en
Since $f_{n\,2}=(f_{1\,2})^{1/n}$, it easy to verify that these double  poles
at $r_i$ are present, with the same coefficient, for any value of $n$. 
Therefore  the regular zeros of $f_{n\,2}'$, where the mapping
$w\to f_{n\,2}$ is not invertible, behave like  (fake) primary operators of 
complex scaling dimensions $\delta_n=\bar\delta_n=-\frac c{16}$. Hence 
 $\{f_{n\,2},w\}$ has unwanted  singularities besides those associated to the branch points.

On the contrary, in the case of a single interval   it is straightforward 
to write down the inverse function $w(z)=f^{-1}_{n,1}(z)$ which solves
the ordinary differential equation (ODE)
\eq
w'^{~n}\propto 
(w-u)^{n-1}(w-v)^{n+1}~~,
\label{t1}
\en 
showing that the only points where this conformal mapping is not invertible
are just the branch points $u$ and $v$.

Having proved that when $N>1$ (\ref{fwrong}) is not a solution of 
(\ref{twrong}), the question  arises as to whether  there is any 
solution of such an equation. 
A crucial observation is that
the $n-$sheeted Riemann surface $\R_{n,N}$ with $N>1$  is topologically 
inequivalent to  the complex
plane $\C$. {\sl A fortiori} there is no way to conformally map 
  $\R_{n,N>1}$ onto $\C$. As a consequence, one cannot longer use the 
main physical motivation outlined at the beginning of this Section, namely
 that the vanishing of $\bra T \ket_\C$ leads to  express 
$\bra T(w)\ket_{\R_{n,N>1}}$ as a Schwarzian derivative. However this argument 
does not exclude {\sl a priori} that $\bra T(w)\ket_{\R_{n,N>1}}$ is a 
Schwarzian derivative for some other reason. Thus we
can try to find possible solutions of (\ref{twrong}).

From a mathematical point of view the problem can be reformulated as follows:
find a function $z(w)$ such that \emph{i)} its Schwarzian derivative 
$\{z,w\}$ is singular only at the branch points $u_1,u_2,\dots, u_{2N}$ 
of $R_{n,N}$; \emph{ii)}
these singularities are double poles; \emph{iii)} the coefficient\footnote{ The last condition  implies 
that the scaling dimension of the associated primary field is $\Delta_n$.} of 
$(w-u_i)^{-2}$  is $\frac{1-1/n^2}{2}$ . We shall see that this 
problem admits very few solutions which can be explicitly constructed.
  
As a starting point of this analysis, notice that it is always 
possible to map  a connected domain of $\C$ (with some boundary identified) 
onto $R_{n,N}$.  
For instance $\R_{2,2}$, the double cover of the plane $w$ branched over 
$u_1,u_2,u_3,u_4$, is conformally equivalent to a torus, represented by a 
parallelogram $\D$ of the $z-$plane with opposite sides identified. The ODE
(\ref{t1}) is now replaced by
\eq
w'^{~2}\propto (w-u_1)(w-u_2)(w-u_3)(w-u_4)~~.
\label{t2}
\en 
Its general solution may be brought into the form
\eq
w(z)\propto\frac1{\wp(z)}+u_1~~,
\label{weie}
\en
where $\wp(z)$ is the elliptic $\wp-$ function of 
Weierstrass associated with the mentioned parallelogram. Since the 
singularities of $\{z,w\}$ coincide with the zeros of $w'$ the first condition 
is fulfilled. An explicit, straightforward, calculation yields
\eq
\{z,w\}=\sum_{i=1}^4\left[\frac38\frac1{(w-u_i)^2}-
\frac14\frac1{w-u_i}\sum_{j\not=i}\frac1{u_i-u_j}\right]~,
\label{tr22}
\en
which is the sought after behaviour. It shows, once again, that in  
the $\R_{2,2}$ case the solution of (\ref{twrong}) is not (\ref{fwrong}), 
but the inverse of the 
function (\ref{weie}) \footnote{This inverse function may be expressed in
 terms of an elliptic integral of the first kind.}.    

One may envisage a straightforward extension of (\ref{t1}) and (\ref{t2}) 
by looking for solutions of  the equation
\eq
w'^{~n}\propto{\sf P}(w)~~,
\label{tp}
\en
where ${\sf P}(w)$ is zero or singular only on the branch points 
$u_j,v_j$  $(j=1,\dots,N)$.  It may be worth observing that  
evaluating $\{z,w\}$ does not require the actual knowledge of the 
solution of (\ref{tp}): using  the identity $\{z,w\}=-\{w,z\}/w'^{~2}$ 
we obtain
\eq
\{z,w\}=\frac{(2n-1)\,\dot{\sf P}^{~2}- 2n\,
\ddot{\sf P}\,{\sf P}}
{2n^2\,{\sf P}^{~2}}~~,
\label{ppp}
\en
where $\dot{\sf P}=\frac{ d\,{\sf P} }{d\,w}$ and
$\ddot{\sf P}=\frac{ d\,\dot{\sf P} }{d\,w}$.

Contrarily to naive expectations, there are very strong restrictions on the 
possible form of $ {\sf P}(w)$ if we are to exclude branch points associated 
to primary fields with negative scaling dimensions. First, ${\sf P}(w)$ must
 be a polynomial of degree $2n$ at most. However, if the degree is less 
than $2n$ there is a branch point at 
infinity on the $n-$sheeted Riemann surface, thus we assume that 
${\sf P}(w)$ is exactly of  degree $2n$ in $w$, but this is not yet 
sufficient. It is worth noting that the absence of negative scaling dimensions
coincide with the requirement that there are no movable singularities
\cite{ince}, i.e. singularities of the solution $w(z)$ which move in 
the $z-$plane as the initial values are varied.
 
If we demand either the absence of movable singularities 
or the positivity of the scaling dimensions of the involved twist fields, it turns out that there are 
essentially only six types of allowed ODE of the kind
(\ref{tp}) \footnote{See for instance \cite{ince}, p.312ff.}. 
Two of them are (\ref{t1}) and (\ref{t2}). A third type is simply
a limit case of (\ref{t2}) when $u_1=u_2$. The other three types are 
\ear
w'^{~3}\propto&(w-u_1)^2(w-u_2)^2(w-u_3)^2~,\\
w'^{~4}\propto&(w-u_1)^3(w-u_2)^3(w-u_3)^2~,\\
w'^{~6}\propto&(w-u_1)^5(w-u_2)^4(w-u_3)^3~.
\rae 
They describe multi-sheeted Riemann surfaces  branched over three points 
as depicted in Figure \ref{Figure:0}. In particular the first, 
once inserted in (\ref{ppp}) 
and combined with the conformal Ward identities, leads to (\ref{three}) 
with $n=3$.  
The latter two correspond to $n=4$ and $n=6$ and could 
therefore be used for hints to extend the OPE (\ref{ope}) to even $n$.

Comparison of (\ref{tr22}) with the conformal Ward identity associated
to $\bra T(w)_{\R_{2,2}}$ suggests
\eq
\tr\rho^2(u_1,u_2,u_3,u_4)\propto
\prod_{j<k\le 4}\frac1{\vert u_j-u_k\vert^{c/12}}~.
\label{r22}
\en
 This expression has the expected symmetry properties under the permutations of the branch points; note also that rescaling all the coordinates by 
a common scale factor 
$u_i\to\lambda\,u_i$ yields $\tr\rho^2(\lambda u_i)=\lambda^{-c/2}\,\rho^2(u_i)$,
which is the expected scaling property of the square correlator of four 
twist fields (see also (\ref{rescaling})). 
However Eq. (\ref{r22}) cannot be trusted because it implies, as it would be simple to show, the unwanted identity $\bra T(z)\ket_{\D}\equiv0$, where $\D$ is the parallelogram of 
the $z-$plane conformally equivalent to $\R_{2,2}$. 

Summing up, we have shown that  Eq. (\ref{twrong}) has no other solutions 
besides the two-point and some three-point correlation functions. Thus the problem of finding the entanglement entropy of disjoint intervals remains open.  

\setcounter{equation}{0}
\section{$\tr \rho_A^n$ as a vacuum expectation value}
\label{rho}
In this section we explicitly write $\tr \rho_A^n$ in whatever quantum system 
as the vacuum expectation value of a suitable observable $\os$, defined on a 
larger system composed of $n$ 
{\sl decoupled} copies of the original system. This method is particularly well 
suited for numerical simulations because a single numerical experiment 
directly yields the value of $\tr \rho_A^n$. 

The partition function $Z=\tr e^{-\beta H}$ of  our 
$d-$dimensional quantum system at inverse temperature $\beta$ can be computed 
  in a standard way  by doing the Euclidean  functional integral in a 
$d+1$-dimensional  hyper-cubic lattice $\Lambda=\{\vec{x},\tau\}$  $(x_i,\tau\in\Z)$ over fields $\phi(x)\equiv
\phi(\vec{x},\tau)$ periodic under $\tau\to\tau+\beta$. Therefore the system 
composed of $n$ independent replicas of the original system is described by 
the $n-$th power of $Z$: 
\eq
Z^n=\int\prod_{k=1}^n\D[\phi_k] e^{-\sum_{k=1}^nS[\phi_k]}~,
\label{nz}
\en   
where $\phi_k$ is a field configuration associated with the $k-$th
replica and $S[\phi]$ is the Euclidean action.
Though the method has a much wider applicability, we assume, for the sake of simplicity, that the fields $\phi_k$ are associated with the nodes of the lattice and 
that $S$ is the sum of contributions of the nodes and the links of the lattice
$\Lambda$
\eq
S[\phi_k]=\sum_{x\in\Lambda}V(\phi_k(x))+\sum_{\bra x\,y\ket}
F(\phi_k(x),\phi_k(y))~,
\label{naction}
\en 
with $\bra x\,y\ket$ ranging over pairs of adjacent nodes in $\Lambda$; 
 $V$ and $F$ are arbitrary functions.
\begin{figure}[tb]
\begin{center}
\mbox{~\epsfig{file=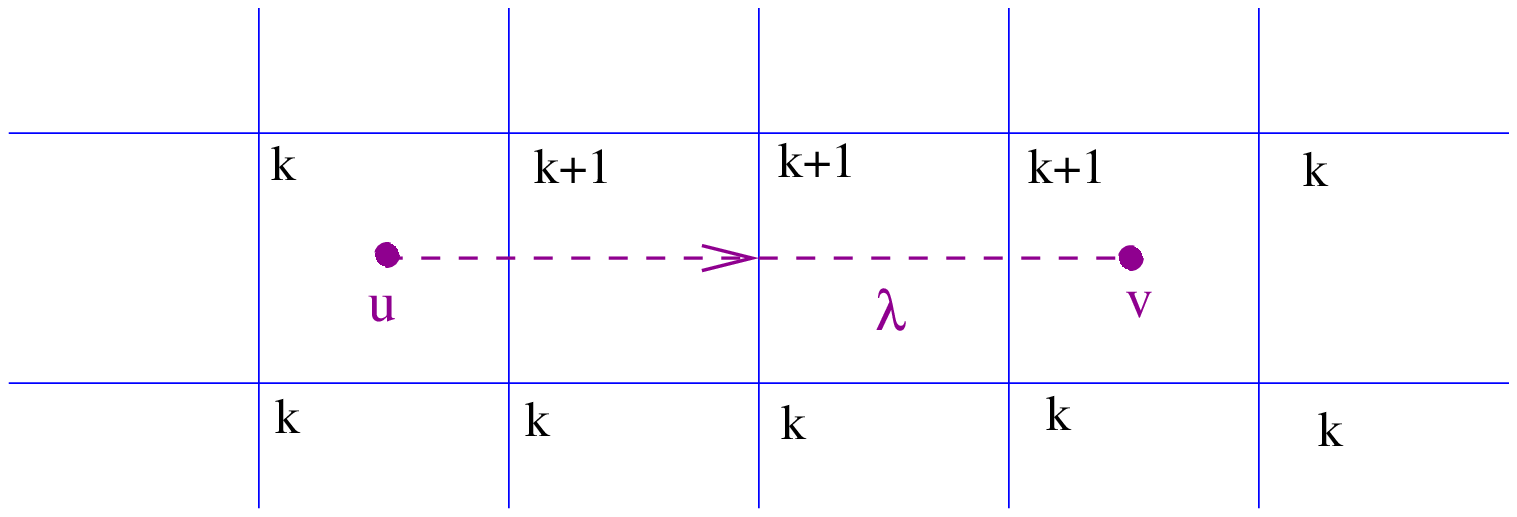,width=8.cm}}
\vskip .2 cm
\caption{The branch points in a square lattice are located on the sites of the dual lattice.  The links intersecting the cut $\lambda$ connect two 
consecutive planar lattices labelled by $k$ 
and $k+1~ ({\rm mod}\, n)$. }
\label{Figure:1}
\end{center}
\end{figure}
   
To define the coupled action $S_A^{(n)}[\phi_1,\phi_2\dots,\phi_n]$, let us 
begin with a two-dimensional system defined in a square lattice. The subsystem 
$A$ consists of one (or more) segments joining pairs of nodes in the dual 
lattice $\tilde\Lambda$. 
We now connect these pairs of nodes  with a line $\lambda$, not 
necessarily coinciding with the segments of $A$, and replace each link 
intersecting this line  with a link connecting two 
consecutive sheets, $k$ and $k+1$, in a cyclical order 
(see Figure \ref{Figure:1}). In this way we obtain  a discretized 
version of a $n-$sheeted Riemann surface where the two points on the dual 
lattice are the branch points and the line $\lambda$ is the cut. 
The associated coupled action is
\eq
S_A^{(n)}=
\sum_{k=1}^n\left[\sum_{x\in \Lambda}V(\phi_k)
+\sum_{\bra x\,y\ket\not\in\lambda}
F(\phi_k(x),\phi_k(y))+
\sum_{\bra x\,y\ket\in\lambda}
 F(\phi_k(x),\phi_{k+1\,({\rm mod}\,n)}(y))\right]~.
\en

The generalisation to higher dimensions is almost obvious: in a 
 $2+1$ dimensional system the two (or more) branch points are replaced by 
one (or more) closed paths $\gamma$ on the dual lattice and $\lambda$ 
is replaced by a surface $\Sigma$ whose boundary is $\gamma$, and so on. 

The observable we are interested in is 
\eq
\os=e^{-\left(S_A^{(n)}[\phi_1,\phi_2,\dots,\phi_n]-\sum_{k=1}^nS[\phi_k]
\right)}~.
\en
Its vacuum expectation value in the system of $n$ independent copies of the
original system is
\eq
\bra\os\ket_n=\frac{\int\prod_{k=1}^n\D\phi_k\,\os\,e^{-S[\phi_k]}}{Z^n}
=\frac{\int\prod_{k=1}^n\D\phi_k\,e^{-S_A^{(n)}}}{Z^n}
=\frac{Z_n(A)}{Z^n}=\tr\rho_A^n~.
\label{obs}
\en
$Z_n(A)$ is the partition function of the system in the $n-$sheeted 
Riemann surface or of its multi-dimensional generalisations. 
The reduced density matrix $\rho_A$ is normalised as in \cite{cc}.

In numerical simulations this observable can be evaluated with great accuracy,
because only the links intersecting the cut $\lambda$ (or its multi-dimensional generalisation) contribute to it. Actually, it is not necessary to simulate $n$ independent systems: simulating a single system and taking, for each 
measurement of $\os$, $n$ statistically independent configurations is 
enough. Alternatively, as a consistency check of the algorithm, one may simulate the system on a $n-$sheeted Riemann surface (or its multi-dimensional 
generalisations)
and verify that in such a case one should obtain
\eq 
\bra \os^{-1}\ket_\R=\frac{Z^n}{Z_n(A)}~.
\en
\subsection{A useful symmetry}
\label{usy}
A system composed of $n$ decoupled copies of the same $d-$dimensional quantum 
system has an interesting invariance which is something more than 
the standard replica symmetry considered in disordered systems.
Let us assume that the system in its Euclidean description is defined on a
stack $\{\Lambda_k~,k=1,\dots n\}$ of $n$ copies of a $d+1$ dimensional 
hyper-cubic lattice. Each node of the stack is characterised by 
$d+2$ integral coordinates $(x_1,x_2,\dots,x_{d+1},k)$ with $1\le k\le n$. 
Assume furthermore an imaginary time coordinate $x_{d+1}$ running 
form $-\infty$ to $\infty$ (i.e. zero temperature) and define the 
following transformation  
\eq
x_i\to x_i ~(i=1,2,\dots d+1)~~;~~~k\to
\cases {k  & if $x_{d+1}\le \alpha$\cr
k-1 ~({\rm mod}~n) & {if $x_{d+1}> \alpha$}
}
\label{transfid}
\en    
where $\alpha$ is an arbitrary real number. This transformation may be 
thought of  as a simple relabelling of nodes of the stack: after such a 
transformation the system consists always of $n$ disjoint lattices, with exactly the same geometric structure as before the transformation; the only 
difference is that the label $k$ is no longer constant along a
given copy. As a consequence, the partition function $Z^n$ of the composed 
system is obviously invariant under such a transformation. Note however that
it is not a symmetry of the classical action $\sum_k S[\phi_k]$: one has to 
perform the functional integration in order to implement this invariance; 
therefore in a sense it can be considered to be a quantum symmetry. 
The above considerations can be generalised in  an obvious way to the 
coupled system  described by the partition function $Z_n(A)$.

\begin{figure}[tb]
\begin{center}
\mbox{~\epsfig{file=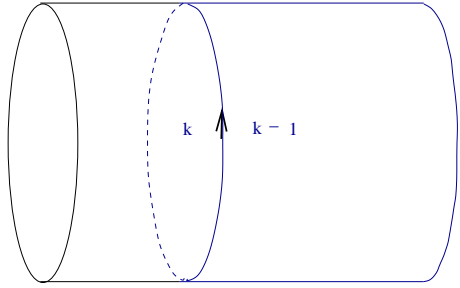,width=6.cm}}
\vskip .2 cm
\caption{The $k$ labels of the nodes of the stack of $n$ cylinders 
after the transformation (\ref{transfid}). }
\label{Figure:1a}
\end{center}
\end{figure}

\begin{figure}[tb]
\begin{center}
\mbox{~\epsfig{file=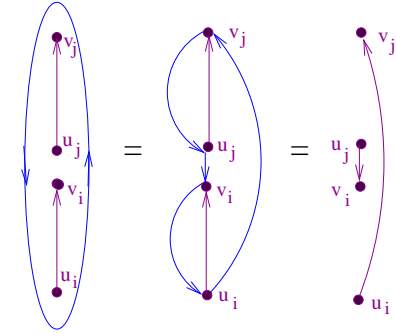,width=5.cm}}
\vskip .2 cm
\caption{A graphical proof of the identity (\ref{symuv}). }
\label{Figure:1c}
\end{center}
\end{figure}

To make the discussion concrete and explicit, we specialise to the case where 
the field theory in question is defined on an infinite cylinder of 
circumference $L$ (see Figure \ref{Figure:1a}). The transformation just 
defined could be also generated by  the $n-$sheeted Riemann surface 
described in Figure \ref{Figure:1}
by moving the two end points in opposite directions in such a way that 
the cut $\lambda$ does wrap  around the cylinder. In the limit where the end 
points touch each other, they annihilate and disappear, and  the 
$n-$ sheeted Riemann surface becomes a stack of $n$ disjoint cylinders.
Thus, the symmetry generated by the transformation (\ref{transfid}) can be 
viewed as the invariance of the system under the addition (or the removal) 
of closed cuts wrapped around the cylinder. We can enlarge the game by 
also considering cuts associated with homotopically trivial loops, which can be 
used for instance to show that the cut $\lambda$ connecting two branch points 
my undergo continuous deformations, or to prove identities of the kind
\eq
\tr\rho^n_{A=\{(u_i,v_i)\,(u_j,v_j)\dots\}}\equiv
\tr\rho^n_{A=\{(u_i,v_j)\,(u_j,v_i)\dots\}}~.
\label{symuv}
\en
 This point is illustrated in Figure \ref{Figure:1c}. In  words, 
these identities show that the cuts or intervals associated with the 
branch points $u_j$ and $v_j$ 
are in no way distinguished lines on the surface: their introduction 
has a similar role as the choice of a reference frame on the surface. 
Note that the essential physics responsible for this symmetry is not some special 
effect of CFT, but rather a general property of the functional
integration - a basic tenet of generic quantum field theories. 

If we combined together open and closed cuts associated with arbitrary permutations of replicas, a very rich structure would  emerge. However it goes beyond the 
scope of this paper. In the following, we consider for convenience only cyclical or anti-cyclical permutations of the replicas. 

\begin{figure}[tb]
\begin{center}
\mbox{~\epsfig{file=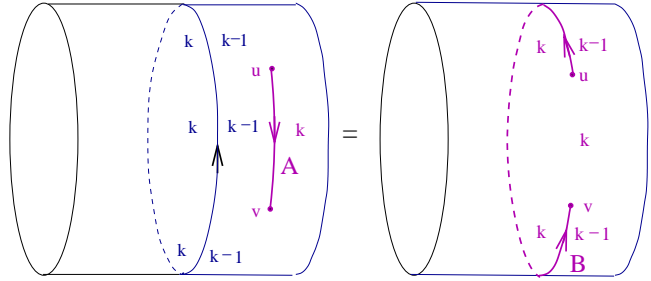,width=8.5cm}}
\vskip .2 cm
\caption{A graphical proof of the equality $\tr\rho_A^n=\tr\rho_B^n$. }
\label{Figure:1b}
\end{center}
\end{figure}
Two points $u$ and $v$ on a cylinder can be connected by a cut $\lambda$ in two topologically inequivalent ways  (see Figure \ref{Figure:1b}). These 
two ways correspond to  the two complementary subsystems $A$ and $B$ of the 
whole quantum system under study. Now if we combine an open cut associated to 
the cyclic permutation $k\to k+1$ with a closed  cut associated with the anti-cyclic permutation $k\to k-1$, like in Figure \ref{Figure:1b}, the mentioned 
symmetry  evidently interchanges the role of the two subsystems $A$ and $B$ 
and yields at once the fundamental identity
\eq
\tr\rho_A^n=\tr\rho_B^n~,\forall\,n~.
\label{rhoAB}
\en  
It is known that this relation, as well as its ensuing consequence 
(\ref{eentropy}), is valid only if the whole system $A\cup B$ is in a 
pure state. This equality  is violated at finite temperature because
a thermal state is in a mixed state, of course. As a consequence we can 
infer that the mentioned symmetry should break down at finite $T$. 
For, since the quantum system  
corresponds to a two-dimensional  Euclidean theory with a compactified 
periodic imaginary time with period $\beta=1/T$, the infinite cylinder reduces 
to a torus of size $L\times \beta$, and the 
transformation (\ref{transfid}) is no longer a symmetry of the system, as it 
transforms $n$ disjoint tori into a single torus of size $L\times n\beta$.

In our numerical experiments we used the vanishing of the difference 
$\tr\rho_A^n-\tr\rho_B^n$ as a sort of thermostat to keep the temperature of the 
simulated system low enough.  
\setcounter{equation}{0}
\section{Entanglement in  Q state Potts Models}
\label{Potts}
One of the simplest and more studied models of statistical mechanics
is the $Q-$state Potts model \cite{potts,wu}, which is the basic system
symmetric under the permutation group of $Q$ elements. It can be
 defined by associating with each node $x$ of an arbitrary lattice $\Lambda$ 
 the field or spin variable $\phi_x=1,2,\dots,Q$ . Its  
partition function at the coupling  $J$ is taken to be
\eq
Z_Q=\sum_{\{\phi\}}e^{-S_Q[\phi]}
\label{zpotts}
\en
where
$ S_Q[\phi] = -\sum_{\bra xy \ket} J\,\delta_{\phi_x \phi_y}$
 with $\bra xy \ket$ ranging over pairs of adjacent
nodes on $\Lambda$.

In a two-dimensional lattice this system has a typical 
order-disorder transition which is continuous in the range $0\le Q\le 4$.
Contrary to the naive expectation, the clusters made of adjacent sites 
with aligned spins do not play an important role at criticality. A different 
definition of cluster was proposed \cite{ck}. These clusters are
defined  as adjacent sites with the same spin connected by bonds  with
probability $p=1-e^{-J}$. Within such a definition, these clusters
behave correctly at the critical point, in the sense that their radius
and the density of the percolating cluster scale with the correct
critical exponents. 
  
The partition function (\ref{zpotts}) can be rewritten 
in terms of these clusters using the Fortuin Kasteleyn representation 
\cite{kf}:
\eq
Z_Q = \sum_{G \subseteq \Lambda}v^{b(G)} Q^{c(G)},
\label{zqv}
\en
where  $ v=\frac p{1-p}=e^J-1$; the summation is over all 
spanning  subgraphs of $\Lambda$, namely the
subgraphs made with all the nodes of $\Lambda$;  $b(G)$ is the
number of edges of $G$, called {\sl active links}, and $c(G)$ the 
number of connected components or Fortuin-Kasteleyn (FK) clusters.
This formulation of the partition function, sometimes called 
di-chromatic polynomial, enables one to generalise $Q$ from
positive integers to real and complex values. In particular $Q=2$
corresponds to the Ising model and $Q=1$ is the random 
percolation problem. 

The implementation of the general method described in Section \ref{rho} 
 to evaluate $\tr\rho_A^n$ is very simple in this case, even if 
the action  in the FK formulation has no longer the form (\ref{naction}), as
$c(G)$, the number of clusters, is a non-local quantity. 
Each configuration of the $Q$-state Potts model is uniquely characterised 
by the location of the active links. Consider a stack of $n$ configurations 
of this kind, defined on a square lattice. Choose an interval $A$ connecting 
two arbitrary nodes of the dual lattice and sew together the $n$ sheets 
according the rules drawn in 
Figure \ref{Figure:1} so as to form a covering of the $n-$sheeted Riemann surface
with a connected square lattice. Note that the sewing operation does not 
modify the local structure of the lattice nor the location of the active links,
while the number of clusters $c(G)$ may change. Therefore, according to 
(\ref{obs}), we obtain the explicit and simple result  
\eq
\tr \rho_A^n=\bra Q^{c_A-\sum_kc_k}\ket_n~,
\label{Q}
\en
where $c_k$ denotes the number of clusters of the $k$-th replica 
before sewing and $c_A$ is the total number of clusters in the stack 
of $n$ replicas sewn together along $A$; the vacuum expectation value is 
taken with respect the system composed of $n$ decoupled copies of the  
Potts model under study. 

Although so far we have considered for the sake of simplicity 
a two-dimensional system, it is clear that the formula (\ref{Q}) still 
holds true in any space dimension, provided one defines appropriately the 
subsystem $A$.    

A first obvious consequence of (\ref{Q})  is that  
in random percolation, i.e. $Q=1$ Potts model,  $\tr\rho_A^n\equiv1$ for any 
$A$ and any $n$:
as intuitively expected, there is no quantum entanglement in random percolation.
 
Some further remarks are in order. The $Q$-state Potts model on a square 
lattice for continuous $Q$ varying between 0 and 4  undergoes  a second order 
phase transition at $v=\sqrt{Q}$. Its critical behaviour is described by 
a CFT with a conformal anomaly $c$ related to $Q$ by \cite{nie}
\eq
\sqrt{Q}=2\cos\frac\pi{m+1}~~;~~c=1-\frac6{m(m+1)}~.
\label{cQ}
\en
Random percolation at the percolation threshold corresponds to $c=0$, as the 
absence of entanglement combined with (\ref{sine}) requires. 
\begin{figure}
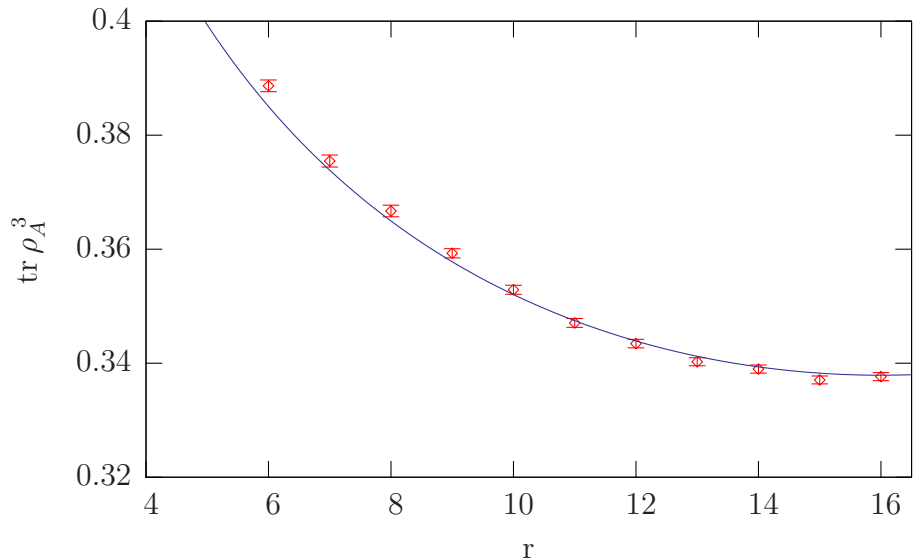

\input isine
\caption{Correlation function of two twist fields 
 in a stack of $n=3$ independent copies of a 2D critical Ising model 
in a square lattice of size $32\times 256$. The data are generated by 
$2\cdot 10^6$ Monte Carlo configurations. 
The solid line is a one-parameter fit to  (\ref{sine}).}
\label{Figure:2}
\end{figure}

For  $Q>1$ there are  very efficient non local cluster Monte Carlo 
algorithms which can be applied  for integer $Q$ \cite{sw,Wolff:1988uh} as
well as for for continuous $Q$ \cite{cm}. They are particularly well suited 
for  accurate estimates of $\tr\rho_A^n$ through the formula (\ref{Q}), as 
at the heart of these algorithms there is a reconstruction of the FK clusters 
of the configurations. 

We simulated in this way a two-dimensional critical Ising model, 
which corresponds  
to a CFT with $c=\frac12$. We considered very elongated lattices 
of size $L\times L'$ with  periodic 
boundary conditions on either side. Typically, the aspect ratios $L'/L$ were
between 4 and 8. This choice amounts to a 
convenient compromise between (\ref{sine}), which is expected to be 
exactly true only 
on an infinitely long cylinder, and  the numerical experiments, which are 
necessarily performed on a finite system. As a criterion to decide whether 
our cylinders were long enough, according with the discussion 
following  (\ref{rhoAB}) we compared the values of $\tr\rho_A^n$ and
$\tr\rho_B^n$ , where $A$ and $B$ are two complementary cuts along the 
circumference of the cylinder, as in Figure \ref{Figure:1b}. When $L'/L=8$  we
found $\tr\rho_A^n\simeq\tr\rho_B^n$ within the statistical errors. The 
behaviour of this quantity  at $n=3$ as a function of the distance of the 
branch points is depicted in Figure \ref{Figure:2}. 
The solid line is a one-parameter fit to (\ref{sine}), where the only 
free parameter is the unknown proportionality constant. We conclude that 
our numerical experiments reproduce very well the expected conformal 
behaviour of the Tsallis entropy, even if the actual size of our lattices 
is not very large. 
\begin{figure}[htb]
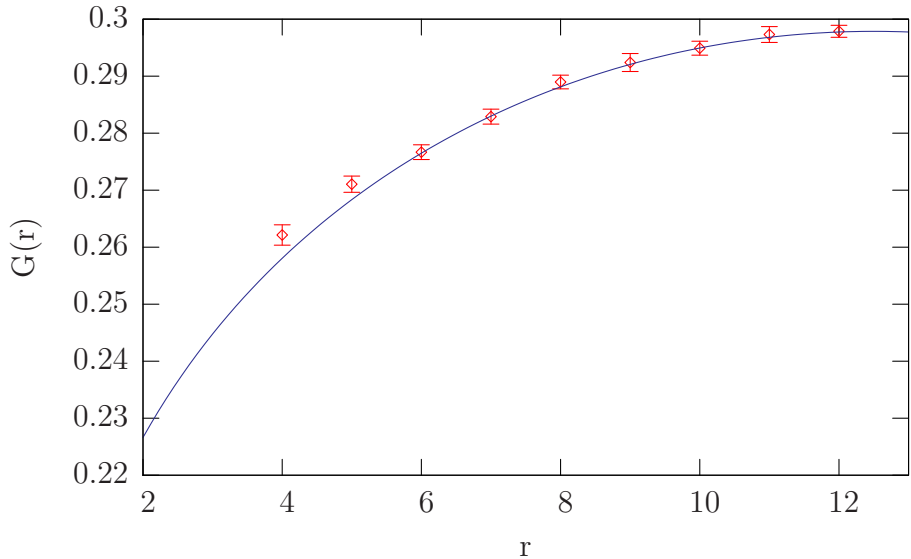

\input sinq
\caption{Correlation function of two twist fields in a stack of $n=2$ 
independent copies of a 2D critical $Q-$state Potts model, with $Q=2-\sqrt{3}$,
 in a square lattice of size $25\times 100$. 
The corresponding CFT has a negative conformal anomaly $(c=-\frac{11}{14})$.
The data come from $2\cdot 10^6$ Monte Carlo configurations generated 
with the algorithm described in \cite{Gliozzi:2007im}. 
The solid line is a one-parameter fit to (\ref{sine}).}
\label{Figure:3}
\end{figure}

When considering the entanglement entropy of critical systems 
corresponding to CFT with negative $c$, an intriguing aspect emerges:
 according to the general expression (\ref{esc}), $S_A$ 
becomes negative. One is led say that the mixed state obtained by 
 tracing over the degrees of freedom of the complement of  
$A$ is in some way more ordered than the vacuum $\vert\Psi\ket$, which has vanishing entropy.
While all that may sound counterintuitive, it has a simple,
testable counterpart in the replica approach: according to (\ref{sine}), 
the correlation function of the branch-point twist fields should 
grow with their distance. One can check it on critical Potts models with
$Q<1$, since they correspond to $c<0$. The non-local cluster algorithms
\cite{sw,Wolff:1988uh,cm} are applicable only for $Q\ge1$, however we can resort to a local algorithm \cite{Gliozzi:2002ub} which can be implemented in an efficient way in the range $0\le Q\le 1$ \cite{dbn,Gliozzi:2007im}.

We performed our numerical experiments at $Q=2-\sqrt{3}$, whose critical 
behaviour is described by the (non-unitary) minimal model $M_{7,12}$, which 
has $c=-\frac{11}{14}$, according to (\ref{cQ}). Also in this case we 
found  good agreement with the predicted behaviour of the two-point twist fields correlator, as  Figure \ref{Figure:3} shows.

\begin{figure}[htb]
\begin{center}
\mbox{~\epsfig{file=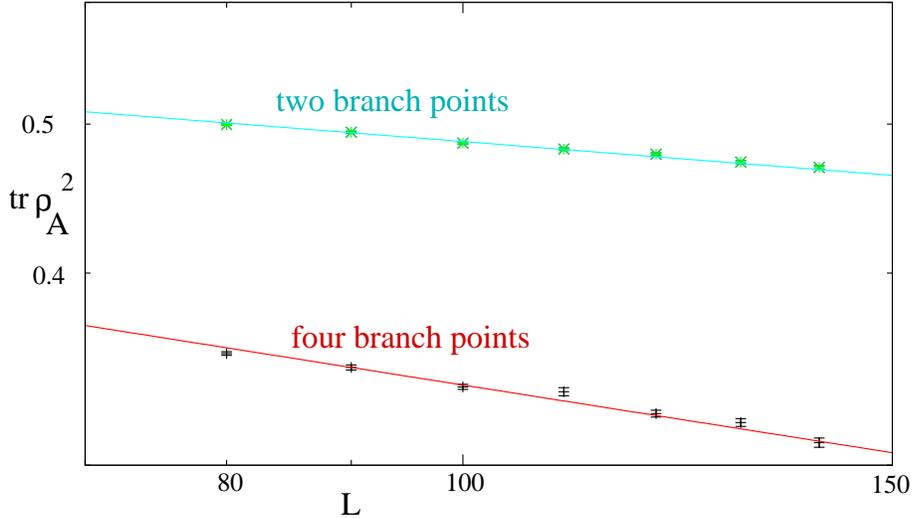,width=12.cm}}
\caption{Comparison of the scaling behaviour of the twist correlator in the 
case of two and four branch points. The $x$ axis is in a log scale and the numerical data lie on straight lines as expected. According to (\ref{rescaling}) 
the slope is exactly
$4n\Delta_nN$, where $2N$ is the number of branch points, while the intercept is the only fitting parameter. In this instance the critical model is the Ising model and $n=2$, therefore $4n\Delta_n=\frac1{8}$. }
\label{Figure:4}
\end{center}
\end{figure}

\begin{figure}[htb]
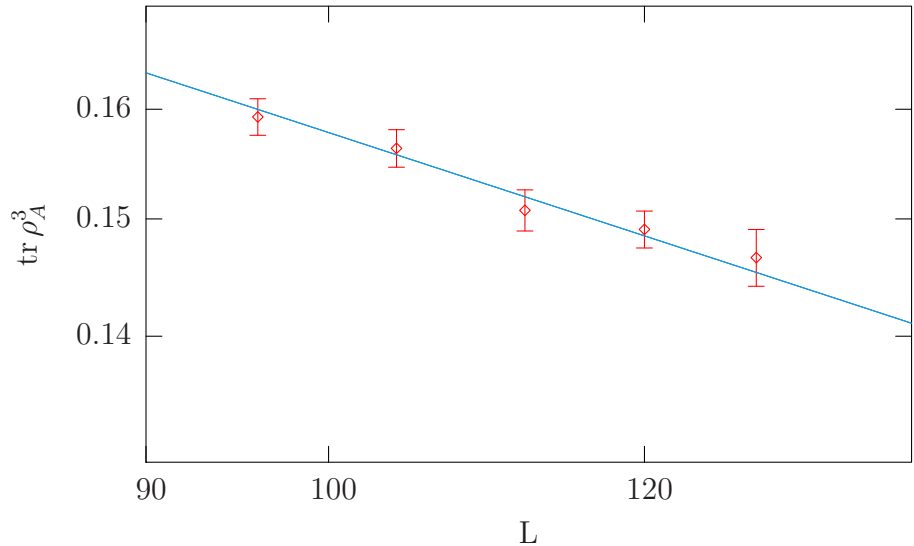

\input scale3_3
\caption{Scaling behaviour of the twist correlator in the 
case of three branch points of a critical Ising model defined on a 
stack of $n=3$ replicas in the 
configuration depicted in Fig.\ref{Figure:0}. 
The $x$ axis is in a log scale and the numerical 
data lie on a straight line as expected. The slope of the solid line is exactly
$6n\Delta_n=\frac13$ as predicted by (\ref{three}). The data of larger lattices are generated by $\sim10^7$ Monte Carlo configurations. }
\label{Figure:5}
\end{figure}

One of the most general and fundamental properties of correlation functions of local operators at criticality is the scaling behaviour under a common rescaling of all the dimensionful quantities. We may use this principle to test 
one of the most important findings of the replica approach to entanglement
\cite{cc}, namely the discovery that  $\tr\rho_A^n$ is proportional to a 
correlation function of twist fields. Although the specific form of 
$\tr\rho_A^n$ in the case in which the whole system is enclosed in a square 
box of side $L$ and the subsystem $A$ consists of $N$ disjoint intervals 
$A\equiv(u_1,v_1),\dots,(u_N,v_N)$ is not actually known, (\ref{rhobp}) 
implies  the following homogeneity property (see Figure \ref{Figure:4})
\eq
\tr\rho^n(\lambda u_1,\lambda v_1,\dots,\lambda u_N,\lambda v_N,\lambda L)=
\lambda^{-4n\Delta_nN}\tr\rho^n( u_1, v_1,\dots, u_N,v_N, L)~,
\label{rescaling}
\en
where $\lambda$ is any positive rescaling factor and $\Delta_n$ is 
the scaling dimension (\ref{delta}) of the $2N$ branch point twist fields. 
We checked this formula in  critical Ising systems  in the cases $N=1$ 
and $N=2$, finding perfect agreement for boxes large enough, as 
Figure \ref{Figure:4} shows: the slopes of the two straight lines 
match precisely with the predicted value $4n\Delta_nN$. We  checked in the 
same way the scaling properties of the three-point function (\ref{three}) 
proposed in the present paper, which corresponds to put $N=\frac32$ in 
the above formula, finding again a good agreement 
(see Figure \ref{Figure:5}), even 
if such a case is  more demanding from a computational point of view.

\begin{figure}[htb]
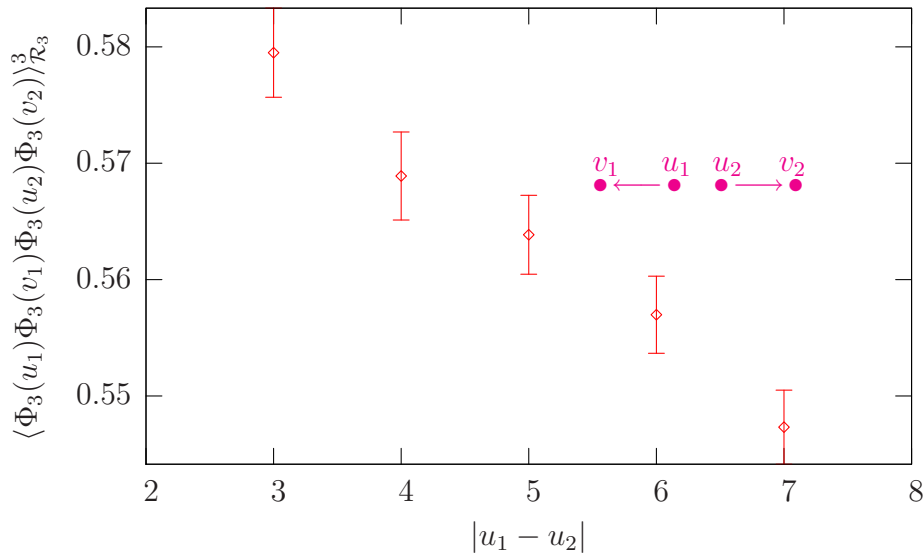

\input uuinter
\caption{In a critical Ising model defined on a stack of $n=3$ replicas, 
the subsystem $A$ is chosen to consist of two collinear segments
$(u_1,v_1)\,(u_2,v_2)$ as indicated in Figure and  the quantity (\ref{r42})
is plotted versus $\vert u_1-u_2\vert$. The size of the lattice is 
$128\times 128$ and $\vert u_1-v_1\vert=\vert u_2-v_2\vert=32$. 
The data are generated by  $2. \, 10^6$ Monte Carlo configurations.}
\label{Figure:6}
\end{figure}
We also performed a numerical experiment to settle whether in the 
limit $\vert u_1-u_2\vert\to0$ the observable
$\tr\rho^n_{A=\{(u_1,v_1)(u_2,v_2)\}}$ vanishes,  as predicted by
(\ref{wrong}), or diverges, as argued in Section \ref{cft}. For this purpose we
considered two collinear intervals, as sketched in Figure \ref{Figure:6}, 
on a stack of 3 replicas. In order to reduce the noise, instead of 
considering as in the other numerical experiments three independent 
Ising systems, we simulated 
the model on a three-sheeted Riemann surface $\R_3$ with a cut in $(u_1,v_1)$, 
thus the measured observable was 
\eq
\bra \Phi_3^-(u_1)\Phi_3^+(v_1)
 \Phi_3^-(u_2)\Phi_3^+(v_2)\ket_{\R_3}\equiv
\frac{\bra \Phi_3^-(u_1)\Phi_3^+(v_1)
 \Phi_3^-(u_2)\Phi_3^+(v_2)\ket_3}
{\bra \Phi_3^-(u_1)\Phi_3^+(v_1)\ket_3}~.
\label{r42}
\en

Figure \ref{Figure:6} shows that this quantity grows as $\vert u_1-u_2\vert$
decreases, even if within this set up it is not possible to accurately measure 
the associated exponent, which according to (\ref{ope}) should be exactly 
half that of a pair of branch points of opposite charge.

\section{Conclusions}
In this paper we discussed both analytically as well as numerically some 
properties of the trace of the $n-$th power of the reduced density 
matrix $\rho_A$ in the special case in which $A$ is a subsystem 
of a quantum system described  by a conformal field 
theory in two dimensions. From the analytical point of view we pointed out 
that when $A$ is composed with more than a single interval the explicit formulae 
proposed in \cite{cc} suffer from some inconsistencies. 
The origin of these  can be traced  to the fact that 
in the derivation of the general formulae certain singularities of the 
Schwarzian derivative have been overlooked. Of course a moot derivation does not imply a  
wrong final result, thus it is well possible that the  formulae for the entanglement 
entropy obtained by analytic continuation to $n\to1$    
are still correct.

From the computational point of view we developed a simple method which allows 
us to directly evaluate  $\tr\rho_A^n$ as a vacuum expectation value. We applied 
 it to two different critical systems in two-dimensional Euclidean lattices,
 corresponding to two different values 
of conformal anomaly $c$, finding perfect agreement with the predicted 
formulae in the case in which $A$ is 
composed  by a single interval in a system of finite size \cite{cc}.

  A distinguishing feature of the present method is that the simulations 
are performed in the unperturbed system: the $n$ replicas of the system 
under  study do not interact with each other nor are sewn together 
along some particular subsystem $A$. The information on entanglement 
is encoded in the observable $\os$ whose evaluation does not perturb the 
system. An obvious advantage of 
such an approach is that one can exploit the same set of configurations to 
obtain information on the entanglement entropy  for a variety of subsystems.

We would also like to remark that the usual method to measure $c$ in lattice 
simulations is very different from that one may infer from the present paper.
The standard method originates from the observation that the first subleading correction to the free energy of a CFT on a long cylinder is universal and proportional to $c$ \cite{car}, however the free energy cannot be directly measured in lattice simulations, thus usually one circumvents this difficulty by measuring and integrating the internal energy, with an inevitable loss of precision. The method described in this paper may be viewed as a novel and powerful tool to 
directly measure $c$: the rescaling property (\ref{rescaling}) of the 
Tsallis entropy tells us that a measurement of $\tr\rho_A^n$ in two 
lattices of different size suffice to fix the  slope of the upper straight 
line drawn in Figure \ref{Figure:4}, and hence the value of $c$.

To conclude, notice that the proposed numerical method can also easily be
applied to critical and non-critical systems in any space-time 
dimensions. For instance, adding one more dimension to the Ising system 
simulated in this paper and applying Kramers 
and Wannier duality \cite{kw}, one may obtain information on the entanglement 
entropy of a confining gauge theory, an issue of growing interest in 
the last years \cite{Nishioka:2006gr,Klebanov:2007ws,
Velytsky:2008rs,Buividovich:2008kq}.   
   
 \section{Aknowledgements}
FG thanks the Galileo Galilei Institute for Theoretical Physics for 
hospitality and the INFN for partial support during the initial phase
of this work. We  also wish to thank R. Tateo for extensive, helpful discussions, 
 P. Calabrese and J. Cardy for fruitful, stimulating correspondence.

\end{document}

%% file: isine.tex
\begin{picture}(0,0)%
\includegraphics{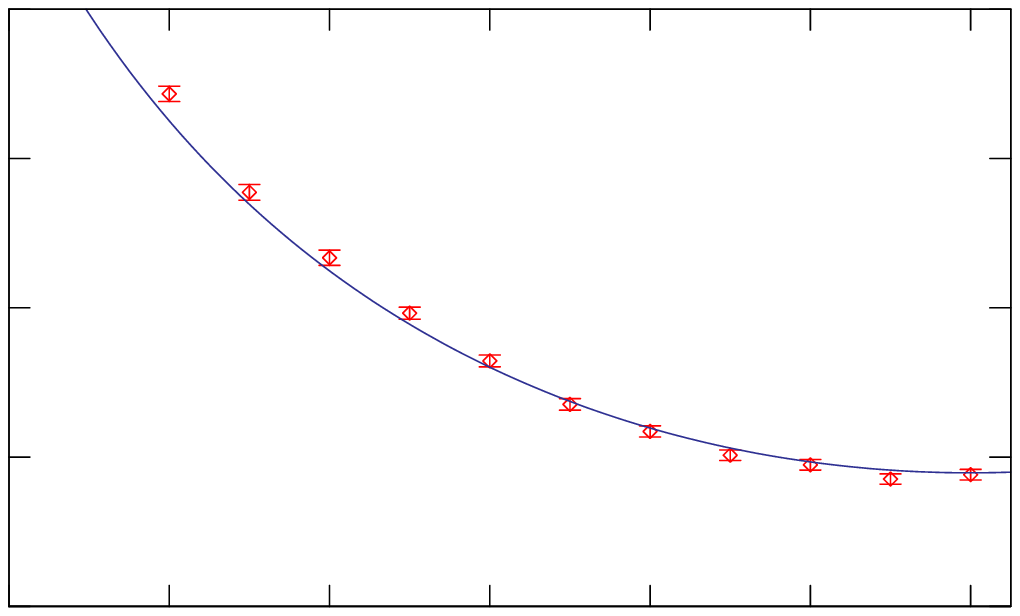}%
\end{picture}%
\begingroup
\setlength{\unitlength}{0.0200bp}%
\begin{picture}(18000,10800)(0,0)%
\put(2475,1650){\makebox(0,0)[r]{\strut{} 0.32}}%
\put(2475,3800){\makebox(0,0)[r]{\strut{} 0.34}}%
\put(2475,5950){\makebox(0,0)[r]{\strut{} 0.36}}%
\put(2475,8100){\makebox(0,0)[r]{\strut{} 0.38}}%
\put(2475,10250){\makebox(0,0)[r]{\strut{} 0.4}}%
\put(2750,1100){\makebox(0,0){\strut{} 4}}%
\put(5058,1100){\makebox(0,0){\strut{} 6}}%
\put(7366,1100){\makebox(0,0){\strut{} 8}}%
\put(9674,1100){\makebox(0,0){\strut{} 10}}%
\put(11982,1100){\makebox(0,0){\strut{} 12}}%
\put(14290,1100){\makebox(0,0){\strut{} 14}}%
\put(16598,1100){\makebox(0,0){\strut{} 16}}%
\put(550,5950){\rotatebox{90}{\makebox(0,0){\strut{}$\tr\rho_{A}^{~3}$}}}%
\put(9962,275){\makebox(0,0){\strut{}r}}%
\end{picture}%
\endgroup
 

%% file: sinq.tex
\begin{picture}(0,0)%
\includegraphics{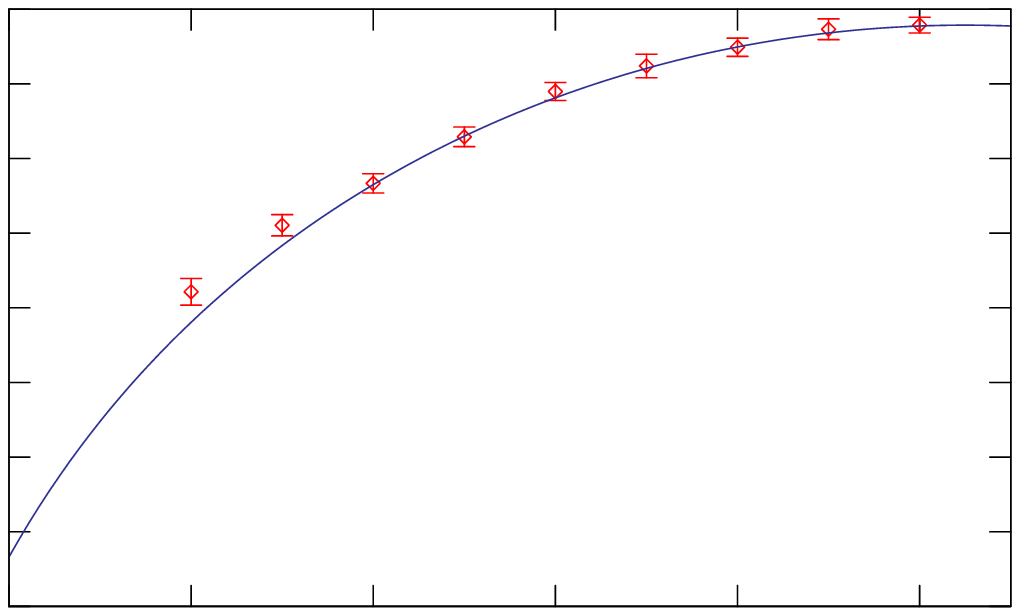}%
\end{picture}%
\begingroup
\setlength{\unitlength}{0.0200bp}%
\begin{picture}(18000,10800)(0,0)%
\put(2475,1650){\makebox(0,0)[r]{\strut{} 0.22}}%
\put(2475,2725){\makebox(0,0)[r]{\strut{} 0.23}}%
\put(2475,3800){\makebox(0,0)[r]{\strut{} 0.24}}%
\put(2475,4875){\makebox(0,0)[r]{\strut{} 0.25}}%
\put(2475,5950){\makebox(0,0)[r]{\strut{} 0.26}}%
\put(2475,7025){\makebox(0,0)[r]{\strut{} 0.27}}%
\put(2475,8100){\makebox(0,0)[r]{\strut{} 0.28}}%
\put(2475,9175){\makebox(0,0)[r]{\strut{} 0.29}}%
\put(2475,10250){\makebox(0,0)[r]{\strut{} 0.3}}%
\put(2750,1100){\makebox(0,0){\strut{} 2}}%
\put(5373,1100){\makebox(0,0){\strut{} 4}}%
\put(7995,1100){\makebox(0,0){\strut{} 6}}%
\put(10618,1100){\makebox(0,0){\strut{} 8}}%
\put(13241,1100){\makebox(0,0){\strut{} 10}}%
\put(15864,1100){\makebox(0,0){\strut{} 12}}%
\put(550,5950){\rotatebox{90}{\makebox(0,0){\strut{}G(r)}}}%
\put(9962,275){\makebox(0,0){\strut{}r}}%
\end{picture}%
\endgroup
 

%% file: scale3_3.tex
\begin{picture}(0,0)%
\includegraphics{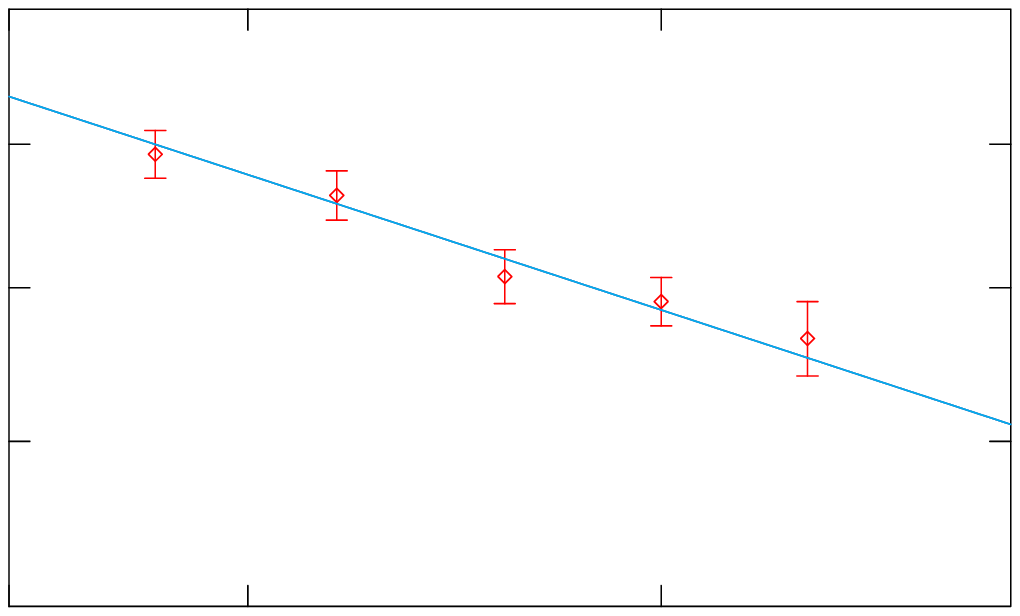}%
\end{picture}%
\setlength{\unitlength}{0.0200bp}%
\begin{picture}(18000,10800)(0,0)%
\put(2475,4026){\makebox(0,0)[r]{\strut{} 0.14}}%
\put(2475,6238){\makebox(0,0)[r]{\strut{} 0.15}}%
\put(2475,8306){\makebox(0,0)[r]{\strut{} 0.16}}%
\put(2750,1100){\makebox(0,0){\strut{} 90}}%
\put(6190,1100){\makebox(0,0){\strut{} 100}}%
\put(12142,1100){\makebox(0,0){\strut{} 120}}%
\put(550,5950){\rotatebox{90}{\makebox(0,0){\strut{}$\tr\rho^3_A$}}}%
\put(9962,275){\makebox(0,0){\strut{}L}}%
\end{picture}%
 

%% file: uuinter.tex
\begin{picture}(0,0)%
\includegraphics{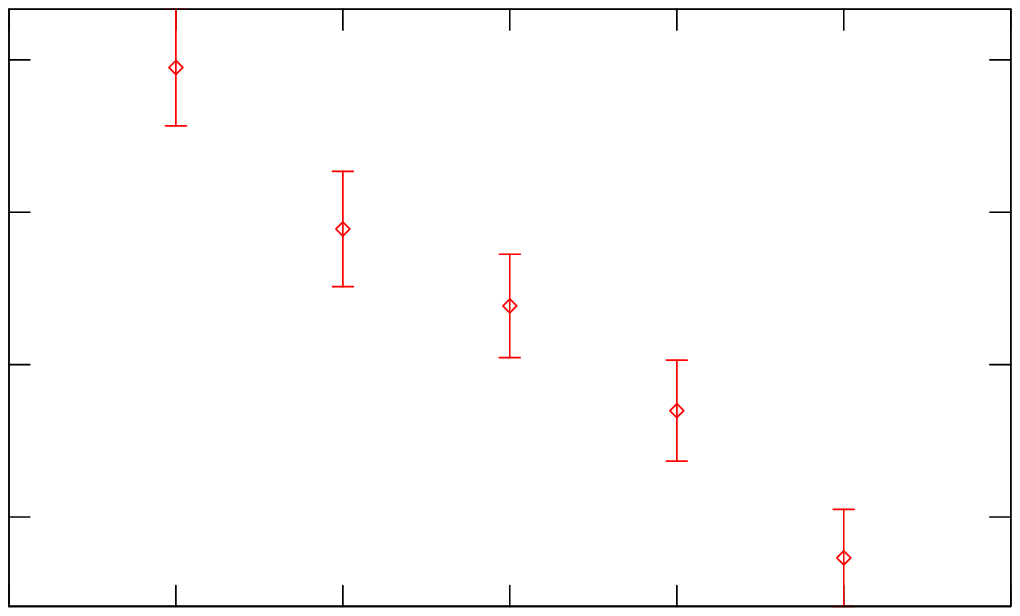}%
\end{picture}%
\begingroup
\setlength{\unitlength}{0.0200bp}%
\begin{picture}(18000,10800)(0,0)%
\put(2475,2937){\makebox(0,0)[r]{\strut{} 0.55}}%
\put(2475,5131){\makebox(0,0)[r]{\strut{} 0.56}}%
\put(2475,7326){\makebox(0,0)[r]{\strut{} 0.57}}%
\put(2475,9520){\makebox(0,0)[r]{\strut{} 0.58}}%
\put(2750,1100){\makebox(0,0){\strut{} 2}}%
\put(5154,1100){\makebox(0,0){\strut{} 3}}%
\put(7558,1100){\makebox(0,0){\strut{} 4}}%
\put(9962,1100){\makebox(0,0){\strut{} 5}}%
\put(12367,1100){\makebox(0,0){\strut{} 6}}%
\put(14771,1100){\makebox(0,0){\strut{} 7}}%
\put(17175,1100){\makebox(0,0){\strut{} 8}}%
\put(550,5950){\rotatebox{90}{\makebox(0,0){\strut{}$\bra \Phi_3(u_1)\Phi_3(v_1)\Phi_3(u_2)\Phi_3(v_2)\ket^3_{\R_3}$}}}%
\put(9962,275){\makebox(0,0){\strut{}$\vert u_1-u_2\vert$}}%
{\color{magenta}
\put(11165,7326){\makebox(0,0)[l]{\strut{}$v_1~~~~ u_1~~u_2~\,~~ v_2$}}%
\put(11165,6920){\makebox(0,0)[l]{\strut{}$\bullet\negthinspace\longleftarrow 
\negthinspace\bullet~~~\bullet
\negthinspace\longrightarrow \negthinspace\bullet$}}}%

\end{picture}%
\endgroup
 